\documentstyle[aps,epsfig]{revtex}

\def\be{\begin{equation}}
\def\ee{\end{equation}}
\def\bea{\begin{eqnarray}}
\def\eea{\end{eqnarray}}

\begin{document}

\title{Measurement of the atmospheric neutrino-induced upgoing muon flux using
MACRO}
\author{{\bf The MACRO Collaboration}\\
M.~Ambrosio$^{12}$, 
R.~Antolini$^{7}$, 
C.~Aramo$^{7,n}$,
G.~Auriemma$^{14,a}$, 
A.~Baldini$^{13}$, 
G.~C.~Barbarino$^{12}$, 
B.~C.~Barish$^{4}$, 
G.~Battistoni$^{6,b}$, 
R.~Bellotti$^{1}$, 
C.~Bemporad $^{13}$
% ...
\\
}
\address{
1. Dipartimento di Fisica dell'Universit\`a di Bari and INFN, 70126 
Bari,  Italy \\
2. Dipartimento di Fisica dell'Universit\`a di Bologna and INFN, 
 40126 Bologna, Italy \\
3. Physics Department, Boston University, Boston, MA 02215, 
USA \\
4. California Institute of Technology, Pasadena, CA 91125, 
USA \\
5. Department of Physics, Drexel University, Philadelphia, 
PA 19104, USA \\
6. Laboratori Nazionali di Frascati dell'INFN, 00044 Frascati (Roma), 
Italy \\
7. Laboratori Nazionali del Gran Sasso dell'INFN, 67010 Assergi 
(L'Aquila),  Italy \\
8. Depts. of Physics and of Astronomy, Indiana University, 
Bloomington, IN 47405, USA \\
9. Dipartimento di Fisica dell'Universit\`a dell'Aquila  and INFN, 
 67100 L'Aquila,  Italy \\
10. Dipartimento di Fisica dell'Universit\`a di Lecce and INFN, 
 73100 Lecce,  Italy \\
11. Department of Physics, University of Michigan, Ann Arbor, 
MI 48109, USA \\        
12. Dipartimento di Fisica dell'Universit\`a di Napoli and INFN, 
 80125 Napoli,  Italy \\        
13. Dipartimento di Fisica dell'Universit\`a di Pisa and INFN, 
56010 Pisa,  Italy \\   
14. Dipartimento di Fisica dell'Universit\`a di Roma ``La Sapienza" and INFN, 
 00185 Roma,   Italy \\         
15. Physics Department, Texas A\&M University, College Station, 
TX 77843, USA \\        
16. Dipartimento di Fisica Sperimentale dell'Universit\`a di Torino and INFN,
 10125 Torino,  Italy \\        
17. Faculty of Sciences, University Mohamed I, B.P. 424 Oujda, Morocco \\
\vspace {0.5cm}
{\footnotesize
$a$ Also Universit\`a della Basilicata, 85100 Potenza,  Italy \\
$b$ Also INFN Milano, 20133 Milano, Italy\\
$c$ Also Istituto TESRE/CNR, 40129 Bologna, Italy \\
$d$ Also Scuola Normale Superiore di Pisa, 56010 Pisa, Italy\\
$e$ Also Dipartimento di Fisica, Universit\`a di Roma Tre, Roma, Italy \\
$f$ Also Universit\`a di Trieste and INFN, 34100 Trieste, 
Italy \\
$g$ Also Dipartimento di Energetica, Universit\`a di Roma, 
 00185 Roma,  Italy \\
$h$ Also Institute for Nuclear Research, Russian Academy
of Science, 117312 Moscow, Russia \\
$i$ Also Institute for Space Sciences, 76900 Bucharest, Romania \\
$l$ Swarthmore College, Swarthmore, PA 19081, USA\\
$m$ RPD, PINSTECH, P.O. Nilore, Islamabad, Pakistan \\
$n$ Also INFN Catania, 95129 Catania, Italy\\
$o$ Also Department of Physics, Pennsylvania State University, 
University Park, PA 16801, USA\\}
}
\date {\today}
\maketitle
\begin {abstract}
We present a measurement of the flux of neutrino-induced upgoing muons
($<$E$_\nu\!> \sim$ 100 GeV) using
the MACRO detector. The ratio of the number of observed to expected events
integrated over all zenith angles is 
0.74 $\pm0.036$(stat) $\pm0.046$(systematic) $\pm0.13$(theoretical).
The observed zenith distribution for $-1.0\le \cos \theta \le -0.1$
does not fit well with the no oscillation expectation, 
giving a maximum probability for $\chi^2$ of 0.1\%.
The acceptance of the detector has been extensively studied
using downgoing muons, independent analyses and Monte-Carlo 
simulations. The other systematic uncertainties cannot be the
source of the discrepancies between the data and expectations.

We have investigated whether the observed number of events and the shape 
of the zenith distribution can be explained by a neutrino 
oscillation hypothesis.  Fitting either 
the flux or zenith distribution
independently yields mixing
parameters
of $\sin^2 2\theta =1.0$ and $\Delta m^2$ of a few times $10^{-3}$ eV$^2$.
However, the observed zenith distribution does 
not fit well with any expectations, 
giving a maximum probability for $\chi^2$ of
5\% for the best oscillation hypothesis, and the 
combined probability for the shape and number of events is 17\%.
We conclude that these data favor a neutrino oscillation hypothesis,
but with unexplained structure in the zenith distribution not 
easily explained by either the statistics or systematics of the experiment.

\vspace {0.3cm}
\begin{center}
{\bf{Submitted to Physics Letters}}
\end{center}
\end {abstract}
            
\pacs{Valid PACS appear here.
{\tt$\backslash$\string pacs\{\}} should always be input,
even if empty.}

\nopagebreak
\twocolumn
\narrowtext

\setcounter{equation}{0}

Over the last decade evidence has been growing for the possibility
of oscillation of atmospheric neutrinos. A first
anomaly was observed 
in the ratio of contained muon neutrino to electron neutrino 
interactions in the IMB \cite {IMB91} and Kamiokande 
\cite{Kamioka92-94} detectors. In addition, the observation of an anomaly
in the multi-GeV atmospheric
neutrino ratio in Kamiokande suggested specific oscillation parameters with
large mixing probability and $\Delta m^2 \approx 10^{-2}$ eV$^2$  
\cite{Kamioka94}. Recent results from Super-Kamiokande have confirmed the
anomaly in the contained event ratio and also show a strong effect in
the zenith angle distribution \cite{SuperK9798} 
suggesting best fit parameters of
$\sin^2 2\theta = 1.0$ and $\Delta m^2$ in the range of a few times
10$^{-3}$ eV$^2$. Also recently, the Soudan 2 detector has
confirmed an anomaly in the $\mu/$e ratio of contained 
events using an iron-based
detector \cite{Soudan9397}. 
Earlier results from the Frejus \cite{Frejus8995} and NUSEX 
\cite{Nusex89} detectors are
consistent with the expected number of contained events 
though with smaller statistics.

The flux of atmospheric muon neutrinos in the energy region
from a few GeV up to hundreds of GeV can be inferred from measurements of
upgoing muons in underground  detectors. If the anomalies in the atmospheric
neutrinos at lower energy are the result of neutrino oscillations, then the
flux of upgoing muons should be affected both in the absolute number of events
observed and in the shape of the zenith angle distribution, with relatively
fewer events observed near the zenith than near the horizontal due to the
longer pathlength of neutrinos near the zenith. Previous measurements of the
upgoing muon flux have been made 
 by the Baksan \cite {Baksan}, Kamiokande \cite {Kam_and_flux}, 
IMB \cite {IMB_and_flux} and Frejus \cite {Frejus8995}
detectors with no claimed discrepancy with expectations from
calculation.

The MACRO detector~\cite{MACRO93} 
provides an excellent tool for the study of upgoing muons.
Its large area (76.6~m~$\times$~12~m~$\times$~9.3~m), 
fine tracking granularity (angular resolution on tracks is between 
0.1$^\circ$ and 1.0$^\circ$), good time resolution (about 500 ps),
symmetric electronics with respect to upgoing
versus downgoing muons and fully-automated analysis permit detailed studies of
the detector acceptance and possible sources of backgrounds to upgoing muons.
In addition, the overburden of the Gran Sasso Laboratory 
(minimum rock overburden of 3150 hg/cm$^2$) is
significantly larger than for the locations of the previous
experiments with the highest statistics on upgoing muons (Baksan and IMB).
This provides additional shielding against possible
sources of background induced by down-going muons.

In our first measurement on upward-going (upgoing) muons,
we reported on a deficit in the
total number of observed upgoing muons with respect to expectations and also an
anomalous zenith angle distribution \cite {MACRO9596}. 
Here, we report on a data set with much higher statistics
\cite {MACRO_Neutrino98}
which retains the same basic features as reported previously.
In addition, an extensive and exhaustive study has been
performed on systematic effects in the analysis and detector acceptance.

The upgoing muon data presented here come from three running 
periods and detector
configurations: the lower half of one supermodule from March 1989 --
November 1991 (1.38 effective live-years), 
the lower half of 6 supermodules from December 1992 --
June 1993 (0.413 effective live-years) and the full detector 
from April 1994 -- December 1997 (2.89 effective live-years).
Results from the first two periods have already been
published~\cite{MACRO9596}.

The sign of the direction that muons travel through MACRO is determined from the
time-of-flight between at least two different layers of scintillator counters
combined with the path length of a track reconstructed in the streamer tubes. 
The measured muon velocity is calculated with the 
convention that muons going down through the detector will be expected to have
1/$\beta$ near +1 while muons going up through the detector will be expected
to have 1/$\beta$ near -1. 
Several cuts are imposed to remove backgrounds caused by radioactivity in near
coincidence with muons and showering events which may result in bad time
reconstruction. The most important cut requires that the
position of a muon hit in each scintillator as determined from the
timing within
the scintillator counter agrees to within $\pm$70 cm of the position
indicated by the streamer tube track.

It has been observed that downgoing muons which pass near or through 
MACRO may produce low-energy, upgoing particles.  These could appear to be
neutrino-induced
upgoing muons if the down-going muon misses the detector \cite {MACROuppi}.
This background has been suppressed by imposing a cut
requiring that each upgoing muon must traverse at least 200 g/cm$^2$ of
material in the bottom half of the detector. 
Finally, a large number
of nearly horizontal ($\cos \theta > -0.1$), but upgoing muons have 
been observed coming from azimuth angles (in local coordinates) from
30$^\circ$-50$^\circ$. This direction corresponds to a cliff in
the mountain where the overburden is insufficient to remove
nearly horizontal, downgoing muons which have scattered in the mountain
and appear as upgoing. We exclude this region from both our
observation and Monte-Carlo calculation of the upgoing events.

\begin{figure}
\begin{center}
\mbox{
\epsfig{file=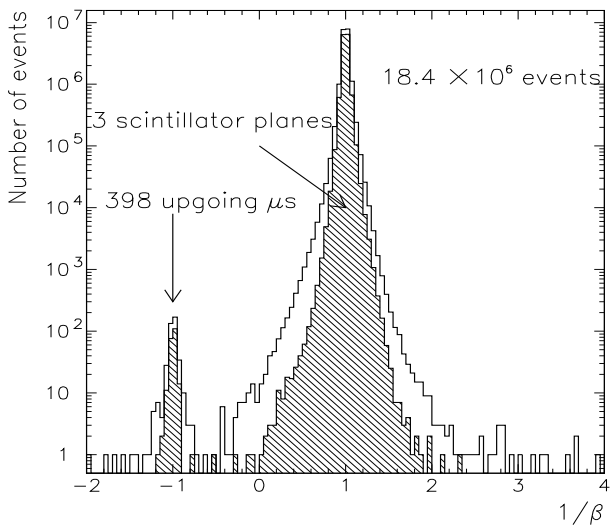,width=9cm,height=8cm}}
\caption {\label{fig:betadis} Distribution of $1/\beta$ for all muons in
  the data set taken with the full detector apparatus. A clear peak of
  upgoing muons is evident centered on $1/\beta=-1$. The widths of the 
  distributions for upgoing and downgoing muons are consistent. The
  shaded part of the distribution is for the subset of events where
  three scintillator layers were hit.}
\end{center}
\end{figure}

Figure \ref {fig:betadis} shows the $1/\beta$ distribution for the
MACRO data from the full detector running (for the older data see the
equivalent figure in Ref. \cite {MACRO9596}). A clear peak of upgoing
muons is evident centered on $1/\beta=-1$. 
There are 398 events in the range $-1.25 < 1/\beta < -0.75$ which we
define as our upgoing muon sample from this data set. We combine
these data with the previously published data (with 4 additional events
due to an updated analysis) for a total of 479 upgoing events.
Based on events outside the upgoing muon peak, we estimate
there are $9 \pm 5$ background events in the total data set.
In addition to these events, we estimate
that there are $8 \pm 3$ events which result from upgoing
charged particles produced by
downgoing muons in the rock near MACRO.
Finally, it is estimated that $11 \pm 4$ events are the
result of interactions of neutrinos in the very bottom layer of MACRO
scintillator. A statistical subtraction from the data is made for
these backgrounds prior to calculation of the flux. Hence, the 
observed number of upward, through-going muons integrated over all 
zenith angles is 451.

A Monte Carlo has been used to estimate the expected number of upgoing
muons.  We use the Bartol neutrino flux~\cite{Agrawal96}, which has a
systematic uncertainty of $\pm14\%$, taking into account the agreement
with measurements of the flux of muons in the atmosphere.
We use the Morfin and Tung
parton set S$_1$~\cite{Morfin91} for calculation of the $\nu N$ 
cross section. These
parton distributions were chosen based on good agreement of the resulting
$\sigma_T$ compared to the world average at $E_\nu = 100$~GeV.
We estimate a systematic error of $\pm9\%$ on the upgoing muon flux due to
uncertainties in $\sigma(\nu N)$, including low-energy
effects~\cite{Lipari94}. The energy loss for muons propagating through
rock is taken from Lohmann {\em et al.}~\cite{Lohmann85}, adjusting
the energy loss for the average composition of rock in the Gran Sasso.
A 5\% systematic uncertainty in the flux of upgoing muons results
from this calculation
due to uncertainties in the rock composition and uncertainties
of muon energy loss. Adding in quadrature all the quoted errors
results in a total systematic uncertainty of 17\% on the expected flux
which is almost uniform with zenith angle. 
The expected upgoing muon fluxes based on
different neutrino fluxes 
\cite {Butkevich89,Mitsui86,Volkova80,Honda,Fluka} are
within 10\% of the value presented here.
The detector has been simulated
using GEANT~\cite{Brun87}, and simulated events are processed in the
same analysis chains as the data. An efficiency factor of 0.97 is
applied to the expected number of events based on various electronic
efficiencies which have been explicitly measured using downgoing muons.

\vspace{-1cm}

\begin{figure}
\begin{center}
\mbox{
\epsfig{file=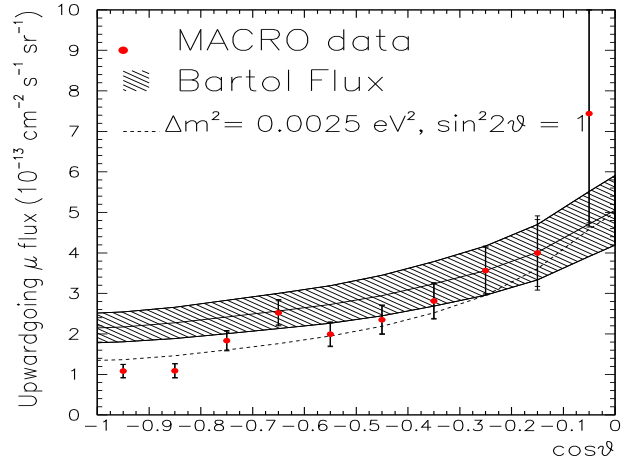,width=9cm,height=7cm}}
\caption {\label{angleflux} Zenith distribution of flux of upgoing
muons with energy
greater than 1 GeV for data and Monte Carlo for the combined MACRO
data. The solid curve shows the expectation for no oscillations and
the shaded region shows the uncertainty in the expectation. The
dashed line shows the prediction for an oscillated flux with
$\sin^2 2 \theta = 1$ and $\Delta m^2 = 0.0025$ eV$^2$.}
\end{center}
\end{figure}

\vspace{-0.5cm}
Care has been taken to ensure a complete simulation of the detector
acceptance in the Monte Carlo and to minimize the systematic
uncertainty in the acceptance.
Comparisons have been made between several different analyses
and acceptance calculations,
including separate electronic and data acquisition systems.
Studies have been made on trigger inefficiencies,
background subtraction, streamer tube efficiencies, and efficiencies
of all data quality cuts. Data distributions over many
different variables (positions of events, azimuth angle, time
distributions, etc.) have been studied and shown to be consistent with
expectations.  
The sum (in quadrature)
of all the systematic errors on the acceptance is $\pm6\%$ for
the total number of events. 
The systematic uncertainty on the acceptance for 
zenith angle bins around the horizon is larger than near the
vertical due to detector 
geometry effects and smaller statistics for downgoing muons.

The number of events expected integrated over all zenith angles is
612, giving a ratio of the observed number of events to the expectation 
of 0.74 $\pm0.036$(stat) $\pm0.046$(systematic) $\pm0.13$(theoretical). 
The probability to obtain a result at least as far from unity as this
is 0.0003
if the Bartol Monte Carlo represents the true parent
flux of neutrinos. However, taking into account the relatively large 
theoretical uncertainty on the flux (mostly on the normalization), the
same probability is 0.14. Hence, there is a low probability that the
Bartol neutrino flux represents the true flux of upgoing neutrinos at
MACRO, but the uncertainty on the normalization of this flux makes it
difficult to conclude (from this test alone)
that new physics, such as neutrino oscillations,
must be responsible for the discrepancy.

Figure~\ref{angleflux} shows the zenith angle distribution of the measured
flux of upgoing muons with energy greater than 1 GeV for all MACRO data
compared to the Monte Carlo expectation for no oscillations (solid
line) and with an oscillated flux with $\sin^2 2 \theta = 1$ and 
$\Delta m^2 = 0.0025$ eV$^2$ (dashed line).
The range for the Monte Carlo expectation 
for the unoscillated flux reflects the  $\pm17\%$ 
systematic uncertainty in that prediction.
The shape of the angular distribution is different than the
expectation giving a $\chi^2 = 26.1$ for 8 degrees of freedom
(probability of 0.001 for a shape at least this different from the
expectation) for the case of no oscillations but with the
number of events in the Monte Carlo normalized to the number in the data.

\vspace{0.5cm}
\begin{figure}
\begin{center}
\mbox{
\epsfig{file=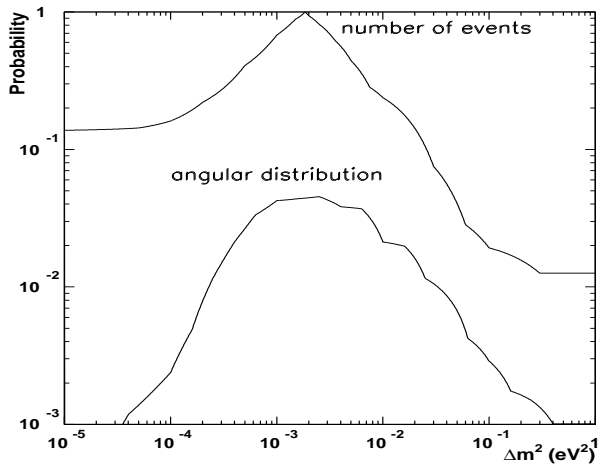,width=9cm,height=7cm}}
\end{center}
\vspace{-2cm}
\caption {\label{fig:dmprob_tau} Probabilities for obtaining the
  observed MACRO results on upgoing muons for $\nu_\mu \rightarrow \nu_\tau$
  oscillations with $\sin^2 2\theta = 1.0$
  and for $\Delta m^2$ as shown. For the number of events, the curve
  shows the probability of observing a number of events which differs
  from the expectation by at least as much as the MACRO data for
  a given value of $\Delta m^2$. For the angular distribution, the
  curve shows the probability to observe a distribution which is at
  least as unlike the expectation based on a $\chi^2$
  comparison of the shape of the data as a function of zenith angle.}
\end{figure}

To test oscillation hypotheses, we calculate the independent probability for
obtaining the number of events observed and the angular distribution
for various oscillation parameters. Figure \ref {fig:dmprob_tau} shows
the probability of obtaining a number of events which differs from the
expectation by at least as much as the MACRO observation for
$\sin^2 2\theta =1.0$ and various $\Delta m^2$ for
$\nu_\mu \rightarrow \nu_\tau$ oscillations. (This is a two-sided 
Gaussian probability.)
The expectation for
$\Delta m^2 = 0.002$ eV$^2$ agrees with the observed number of events.

The probability of $\chi^2$ for obtaining the observed shape of the angular
distribution has been computed as above for oscillation hypotheses and
is also presented in Fig. \ref {fig:dmprob_tau}
for $\nu_\mu \rightarrow \nu_\tau$ oscillations.
The number of events under different flux hypotheses
is always normalized to the observed number of events for this
comparison. A maximum probability of 5\% is obtained for a distribution at
least this different from the expectation for $\nu_\mu \rightarrow
\nu_\tau$ oscillations. This occurs for $\sin^2 2\theta=1.0$ and 
$\Delta m^2=0.0025$ eV$^2$, but the probability is changed little 
within a decade of $\Delta m^2$ around this value. However, it is
notable that the same best value for $\Delta m^2$ is obtained 
independently from both the angular distribution and the number of events.
The somewhat low probability 
for any of these hypotheses is the result of the
relatively low number of events in the region $-1.0<\cos\theta<-0.8$
compared to the number of events in the region $-0.8<\cos\theta<-0.6$. 

\begin{figure}
\begin{center}
\mbox{
\epsfig{file=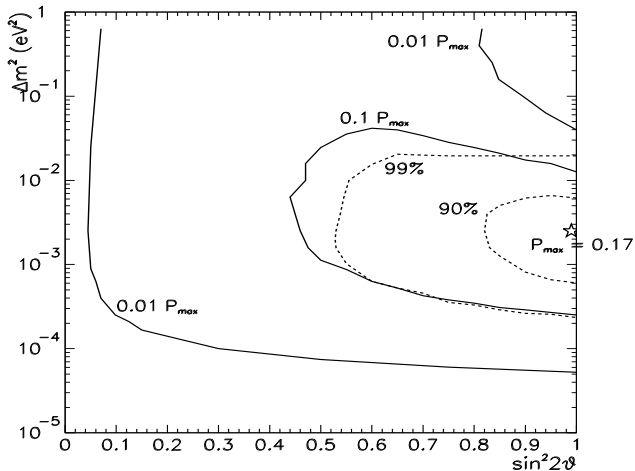,width=8.5cm,height=7cm}}
\end{center}
\caption {\label{fig:exclusion} Probability contours
  for oscillation parameters for $\nu_\mu \rightarrow \nu_\tau$
  oscillations based on the combined probabilities of zenith shape and
  number of events tests. 
   The best-fit point has a probability of 17\% and iso-probability
   contours are shown for 10\% and 1\% of this value (i.e. 1.7\% and
   0.17\%). The dashed lines show exclusion confidence intervals 
   at the 90\% and 99\% levels calculated according to reference 
   \protect\cite {Feldman-Cousins}. 
   Since the best probability is outside the physical region
   the confidence intervals regions are smaller than the one expected
   from the sensitivity of the experiment.
   The ``sensitivity'' contour (not shown) is slightly larger than
   that for 10\% of $P_{\rm max}$.}
\end{figure}

Figure \ref {fig:exclusion} shows probability contours for oscillation
parameters using the combination of probability for
the number of events and $\chi^2$ of the angular distribution.
 The best-fit
point has a probability of 17\%.
The solid lines show the probability contours for 10\% and 1\% of the
best-fit value (i.e. 1.7\% and 0.17\%). The dashed lines show the
exclusion contours at the 90\% and 99\% confidence levels based on
application of the Monte Carlo prescription of reference \cite
{Feldman-Cousins}. The ``sensitivity'' (not shown) is slightly larger than
the curve for 10\% of $P_{\rm max}$. The sensitivity is
the 90\% contour which would result from the preceding prescription if
the data and Monte Carlo happened to be in perfect agreement at the
best-fit point. It should be noted that this prescription for
producing confidence-level intervals assumes that the hypothesis is
correct.

  Possible systematic effects have been studied and shown to be too small to
explain the observed anomalous shape in the
zenith distribution. The detector acceptance is best
understood (from downgoing muons) near the vertical, where the biggest 
deviation compared to the Monte Carlo without oscillations is observed.
The data from all running periods are
consistent in the shape of the zenith distribution.
We have compared the zenith distribution of
down-going muons with a Monte Carlo expectation based on the known
overburden; the two distributions agree well. We have compared the
measured flux of downgoing muons using the same analysis as for the
upgoing muons and find the result is consistent with expectations
(see ref. \cite {MACRO_slant}). The possibility of a
water-filled cavern below MACRO has been studied, although no such caverns
are known to exist in the region of the Gran Sasso. If all of the region
below MACRO were water, a maximum 15\% depletion would be observed in
the flux of upgoing muons. Any realistic water-filled cavern would
result in a depletion of no more than about 5\%. 
For MACRO, we have shown that upgoing charged particles
produced by downgoing muons contribute a background of 2\% of the
total number of upgoing muons \cite {MACROuppi}. This rate could be higher for
experiments located in laboratories with less overburden than the Gran
Sasso.

  It has recently been suggested that oscillations between $\nu_\mu$
and a sterile $\nu$ could qualitatively produce a shape in the zenith
distribution of upgoing muons similar to that observed by MACRO \cite
{Smirnov}. This would result from a matter effect in the center of the Earth.
However, due to suppressed oscillation amplitude, the
current model does not offer a better quantitative
agreement with MACRO data than the $\nu_\mu \rightarrow \nu_\tau$
hypothesis, giving a maximum probability of 2\%.

 In conclusion, we have reported on a measurement of the flux of upgoing muons,
produced by neutrinos (with $<E_\nu\!> \sim 100$ GeV)
originating in atmospheric cosmic-ray showers.
The ratio of the number of observed to expected events
integrated over zenith angles from $-1.0 \le \cos \theta \le -0.1$ is 
0.74 $\pm0.036$(stat) $\pm0.046$(systematic) $\pm0.13$(theoretical).
The observed zenith distribution does not fit well with the expectation, 
giving a maximum probability for $\chi^2$ of only 0.1\%.
The acceptance of the detector has been extensively studied
using downgoing muons, independent analyses, and Monte-Carlo 
simulations. The remaining systematic uncertainties cannot be the
source of the discrepancies between the data and expectations.
We have investigated if the anomaly could be the result of neutrino
oscillations.  Both techniques independently yield mixing parameters 
of $\sin^2 2\theta =1.0$ and $\Delta m^2$ of a few times $10^{-3}$ eV$^2$.
However, the observed zenith distribution does 
not fit well with any expectations, 
giving a maximum probability for $\chi^2$ of only 
5\% for the best oscillation hypothesis.
We conclude that these data favor a neutrino oscillation hypothesis,
but with unexplained structure in the zenith distribution not 
easily explained by either the statistics or systematics of the experiment.

\vspace{-0.7cm}

We are analyzing other topologies of neutrino events.
We will publish shortly the complementary results
from semi-contained events with the neutrino interaction within the
detector ($<E_\nu\!> \sim $ few GeV) and upward going stopping
muons  \cite {MACRO_Neutrino98}.

We gratefully acknowledge the staff of the {\it Laboratori Nazionali
del Gran Sasso} and the invaluable assistance of the technical
staffs of all the participating Institutions.  For generous
financial contributions we thank the U.S.~Department of Energy,
the National Science Foundation, and the Italian {\it Istituto
Nazionale di Fisica Nucleare}, both for direct support and for FAI
grants awarded to non-Italian MACRO collaborators.


\begin{thebibliography}{40}
\bibitem{IMB91}
IMB Collaboration, D. Casper {\it et al.},
Phys. Rev. Lett. {\bf 66}  (1991) 2561;
R. Becker-Szendy {\it et al.}, Phys. Rev. {\bf{D46}} (1992) 3720.
%
\bibitem{Kamioka92-94}
Kamiokande Collaboration, K.S. Hirata {\it et al.}, Phys. Lett. 
{\bf{B205}} (1988) 416; Phys. Lett. {\bf B280} (1992) 2.
%
\bibitem{Kamioka94}
Kamiokande Collaboration, Y. Fukuda {\it et al.}, 
Phys. Lett. {\bf B335} (1994) 237.
%
\bibitem {SuperK9798} Super-Kamiokande Collaboration, 
preprint hep-ex/\-9805006 
``Study of the atmospheric neutrino flux 
in the multi-GeV energy range'',
preprint hep-ex/\-9803006 ``Measurement of a small atmospheric
muon-neutrino/electron-neutrino ratio'' submitted to Phys. Lett. {\bf{B}}.
%
%
\bibitem{Soudan9397}
Soudan 2 Collaboration, 
W.W. M. Allison {\it et al.},
Phys. Lett. {\bf{B391}} (1997) 491;
%
T. Kafka, 5th TAUP Workshop proceedings, Gran Sasso, Italy, 1997.
%
\bibitem{Frejus8995}
Frejus Collaboration, Ch. Berger {\it et al.}, Phys. Lett. {\bf B227} 
(1989) 489;
K. Daum {\it et al.}, Z. Phys. {\bf C66} (1995) 417.
%
%
\bibitem{Nusex89}
NUSEX Collaboration, 
M. Aglietta {\it et al.}, Europhys. Lett. {\bf 8} (1989) 611;
%
23rd ICRC proceedings, Calgary, Vol. {\bf 4} (1993) 446.
%
\bibitem{Baksan} Baksan Collaboration, 
S. Mikheyev, 5th TAUP Workshop proceedings, Gran Sasso, Italy, 1997.
\bibitem{Kam_and_flux} Kamiokande Collaboration, 
Y. Totsuka {\it et al.}, Nucl. Phys. {\bf B31} (Proc. Suppl.)  (1993) 428;
M. Mori {\it et al.}, Phys. Lett. {\bf B270} (1991) 89.
\bibitem{IMB_and_flux} IMB Collaboration, 
R. Becker-Szendy {\it et al.}, Phys. Rev. Lett.
 {\bf 69} (1992) 1010; Nucl. Phys. {\bf{B38}} (Proc. Suppl.) (1995) 331.
\bibitem{MACRO93} 
MACRO Collaboration, S. Ahlen {\it et al.}, Nucl. Instr. and Meth. {\bf A324} 
(1993) 337.
%
\bibitem{MACRO9596} 
MACRO Collaboration, S. Ahlen {\it et al.}, Phys. Lett. {\bf B357} 
(1995) 481; ``Neutrino Induced Upward-going muons with the MACRO Detector'',
Neutrino 96 conference proceedings, Helsinki, 1996; ``The measurement of the 
atmospheric muon neutrino flux using MACRO'', INFN/AE-97/21 and Proc. of the 
25th ICRC, Durban, Vol. {\bf 7} (1997) 41.
%
\bibitem {MACRO_Neutrino98}
MACRO Collaboration, ``Upward-going muons and MACRO'',
presented by F. Ronga at Neutrino 98, Takayama, Japan, 1998
(to be published in conference proceedings).
%
\bibitem{MACROuppi}
MACRO Collaboration, M. Ambrosio {\it et al.}, ``The observation of
upgoing charged particles produced by high energy muons in underground
detectors'', accepted for publication in Astroparticle Physics.
%
\bibitem{Agrawal96}
V. Agrawal, T.K. Gaisser, P. Lipari and T. Stanev, Phys. Rev. {\bf D53}
(1996) 1314.
%
\bibitem{Morfin91}
J. G. Morfin and W. K. Tung , Z. Phys. {\bf C52} (1991) 13.
%
\bibitem{Lipari94}
P. Lipari, M. Lusignoli and F. Sartogo, 
Phys. Rev. Lett. {\bf 74} (1995) 4384. 
%
\bibitem{Lohmann85}
W. Lohmann {\it et al.}, CERN-EP/85-03 (1985).
%
\bibitem{Brun87}
R. Brun {\it et al.}, ``GEANT'', CERN report DD/EE84-1 (1987).
%
\bibitem{Butkevich89}
A.V. Butkevich {\it et al.}, Sov. J. Nucl. Phys. {\bf 50} (1989) 90.
%
\bibitem{Mitsui86}
K. Mitsui {\it et al.}, Nuovo Cimento {\bf 9C} (1986) 995.
%
\bibitem{Volkova80}
L. V. Volkova, Sov. J. Nucl. Phys. {\bf 31} (1980) 1510.
%
\bibitem {Honda}
M. Honda {\it et al.}, Phys. Rev. {\bf D52} (1995) 4985.
%
\bibitem {Fluka}
G. Battistoni {\it et al.},  ``A new calculation of the atmospheric
neutrino flux: The FLUKA approach'', Submitted to Nucl. Phys. {\bf{B}} (1998).
%
\bibitem {Feldman-Cousins}
G. Feldman and R. Cousins, Phys. Rev. {\bf D57} (1998) 3873. 
%
\bibitem {MACRO_slant}
MACRO Collaboration, M. Ambrosio {\it et al.}, Phys. Rev. {\bf D52}
(1995) 3793.
%
\bibitem {Smirnov} Q.Y. Liu and A.Y. Smirnov, preprint IC--97-211, 
hep-ph/9712493.

\end{thebibliography}
\end{document}